\documentclass[prl,aps,superscriptaddress,twocolumn,notitlepage,showpacs]{revtex4-2}
\usepackage{latexsym}
\usepackage{amsmath}
\usepackage{amssymb}
\usepackage{graphicx}
\usepackage{caption}
\usepackage{subfigure}
\usepackage{float}
\usepackage{mathrsfs}
\usepackage{color}
\usepackage{mathrsfs}
\usepackage{txfonts}
\usepackage[justification=centering,
            format=plain]{caption}

\renewcommand{\raggedright}{\leftskip=0pt \rightskip=0pt plus 0cm}
\begin{document}

\title{Quantum Fluctuations and Coherence of a Molecular  Polariton Condensate}

\author{Zhedong Zhang}
\email{zzhan26@cityu.edu.hk}
\affiliation{Department of Physics, City University of Hong Kong, Kowloon, Hong Kong SAR}
\affiliation{City University of Hong Kong, Shenzhen Research Institute, Shenzhen, Guangdong 518057, China}

\author{Shixuan Zhao}
\affiliation{Department of Physics, City University of Hong Kong, Kowloon, Hong Kong SAR}

\author{Dangyuan Lei}
\affiliation{Department of Materials Science and Engineering, City University of Hong Kong, Kowloon, Hong Kong SAR}



\date{\today}

\begin{abstract}
A full quantum theory beyond the mean-field regime is developed for an   exciton polariton condensate, to gain a complete understanding of quantum fluctuations. We find analytical solution for the polariton density matrix, showing the polariton nonlinearity causing fast relaxation correlated with the pump so as to yield the condensation at threshold. Increasing the pump intensity, a nonequilibrium phase transition towards the condensation of lower polaritons emerges, with a statistics transiting from a thermal, through a super-Poissonian and to a nonclassical distribution beyond the understanding at the level of off-diagonal long-range order. The results signify the role of dark states for polariton fluctuations, and lead to a nonclassical counting statistics of emitted photons, which elaborates the role of the key parameters, e.g., pump, detuning and temperature.

\end{abstract}

\maketitle

{\it Introduction}.--The cavity polaritons, formed by strong interaction between excitons and photons, draw much attention in recent years, arising from their rich dynamic and kinetic properties \cite{Coles_NM2014,Ebbesen_AC2012,Spano_PRL2016,Narang_PRL2018,Xiong_NP2020,Schneider_RPP2016,Malpuech_PRL2008}. Compared to atoms and qubits, the exciton polaritons in semiconductors or organic molecules may have a crosstalk with phonon modes undergoing a nonradiative process. This leads to controllable excited-state relaxation possessing multiple timescales and channels, e.g., the interaction between superradiance, subradiance and dark modes that are absent in atomic ensembles \cite{Tavis_PR1968,Spano_PRL1990,Zhang_JPCL2019,Vidal_NJP2015}. Remarkably, it was found that the polaritons may condensate at room temperature, as a reminiscence of the Bose-Einstein condensation of atoms \cite{Kasprzak_Nature2006,Forrest_NP2010,Yamamoto_RMP2010,Manni_PRL2011,Xiong_SA2018,Mahrt_NM2014,Scully_PRL1999,Smith_book2008}. The polariton condensates, however, belong to a different category, due to their nonequilibrium nature under external pump of energy so as to combat the rapid decay of polaritons. Due to the lighter mass of excitons and far-from-equilibrium nature, the polariton condensates can survive even at hundreds of Kelvins, much higher than that for atomic BEC. Such light-induced collective phase is analogous to lasers, and is thus of broad interest in fundamental study and technical applications, e.g., low-cost optoelectronic devices and high-quality light sources \cite{Xiong_ACSNano2018,Xiong_NM2021,Jin_NRM2019,Xu_Nature2020,XiongQ_SA2021,Cohen_NRM2016,Zasedatelev_NP2019}.

Polaritons exist in a variety of systems, such that a large Rabi splitting in the polaritons with organic molecules has been demonstrated recently, owing to their strong polarization \cite{Coles_NM2014,Feist_NC2016,Rubio_PNAS2017,Ebbesen_NC2015}. This is responsible for cooperative light emission \cite{Bellessa_PRL2012,Lidzey_Nature1998}. 
As a result of the delocalization nature, the cooperative motion of exciton polaritons manifests a long-range quantum entanglement over 100s nm. This may create peculiar properties of relaxation dynamics distinct from before. One of the most prominent signatures is the many-particle effects possessing strong correlations and nonlinearity beyond the perturbative understanding for weak couplings \cite{Xiong_SA2021,Mukamel_PNAS2017}. Elaborate experiments demonstrated the nonlinear infrared response of molecules enhanced by vibrational polaritons \cite{Xiong_SA2019,Chen_Science2021}. The collective dynamics involving many molecules against the local disorder thus emerged, highlighting the important role of dark states in the stablization of polaritons \cite{Zhang_JPCL2019}. Polariton condensates show the effects of considerably modifying the electron transfer as well as radiative decay, in the infrared regime \cite{Zhou_arXiv,Spano_JCP2015,Keeling_PT2017}. The dissipative Gross-Pitaevskii equation has been developed for exciton-polariton condensates \cite{Antonelli_Nonlinearity2019,Carusotto_PRL2007,Kivshar_PRB2014}. This is a mean-field description capable of the polariton dynamics much above the critical points. The quantum fluctuations have been well understood in the vicinity of the mean-field regime, known as the Bogoliubov excitations \cite{Yamamoto_NP2008,Ostrovskaya_PRL2021}. These yield a bottleneck in accessing the critical properties where the quantum fluctuations to all orders are significant. The full transition towards the nonequilibrium condensation of polaritons still remains elusive, especially in molecules strongly interacting with light \cite{Herrera_JCP2020,Zhou_CS2018,Feist_PRX2018,Spano_PRA2017}. 

The polariton dynamics was investigated with the density matrix approach from quantum optics \cite{Vidal_NJP2015,Spano_PRA2017,Zhang_PRA2021,Zhang_JCP2018}, manifesting the coherent exciton-photon coupling in a trade-off with the incoherent couplings to other degrees of freedom, i.e., phonons and disorder \cite{Vidal_NJP2015,Zhang_JPCL2019}. This leads to the kinetic theory capable of describing energy and charge transports,  and thermodynamics \cite{Zhou_NC2019}. The fast nonradiative relaxation of excitons, however, produces the nonlinearity that is significant for polariton condensation but has yet been properly integrated. The properties of dark states are thus an open issue, in light of their large density of the states. Most of previous efforts were devoted to the mean-field theory invoking the decorrelation approximation in the equation for density matrix \cite{Haug_PRB2004,Tassone_PRB1997,Keeling_PRL2013,Keeling_PRA2015,Keeling_PRL2018}. The coherence, as a significant fingerprint of polariton-polariton correlations, cannot be obtained thereby. This calls for a completed understanding of polariton fluctuation and nonlinearity. 

\begin{figure*}[t]
 \captionsetup{justification=raggedright,singlelinecheck=false}
\centering
\includegraphics[scale=0.4]{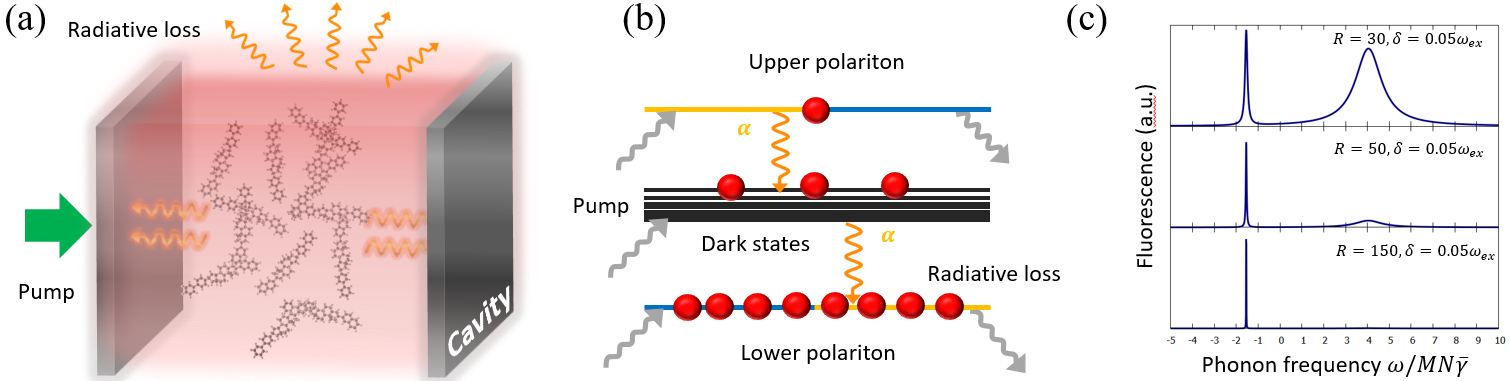}
\caption{(a) Schematic of out-of-equilibrium molecular aggregates in an optical cavity; (b) Energy level diagram of molecular polaritons where, besides radiative loss,  nonradiative couplings between polariton and dark states emerge, as a result of electron-phonon interaction. A strong pump results in a condensation of particles at the lower polariton mode; (c) Fluorescence spectra of the out-of-equilibrium molecular polaritons for various pumping powers; Parameters are the same as Fig.\ref{F2}.}
\label{F1}
\end{figure*}

In this Letter, we develop a full quantum theory for exciton-polariton condensates, to understand the nonequilibrium phase transition in an analogy to the Fr\"ohlich coherence of phonons \cite{Frohlich_IJQC1968}. The density-matrix-based theory is capable of describing the polariton dynamics in three regimes: below threshold, close to threshold and far above threshold. We show the important role of dark states arising at the condensation transition. The density matrix is solved analytically, yielding the number distribution of exciton polaritons whose fluctuations are included to all orders. Notably a non-classical feature is predicted with a strong pumping power, evident by a sub-Poissonian statistics. This is unique for polariton condensates indicating the anti-bunching behavior.


{\it Molecular polariton model}.--We consider a model consisting of $M$ molecular aggregates in an optical cavity, where each includes $N_m$ chromophores. Since the electronic excitations are of the most interest, the chromophore can be modeled by a latticed spinor system interacting with phonons. The Holstein Hamiltonian is $H_{\text{mol}}^{(m)}=\sum_{n=1}^{N_m} h_n^{(m)}$ with
\begin{equation}
    \begin{split}
        h_n^{(m)} =\ & \Delta_m \sigma_n^{(m),+}\sigma_n^{(m),-} + \sum_s\omega_{\text{vib},s}^{(m)} b_{n,s}^{(m),\dagger}b_{n,s}^{(m)}
    \end{split}
\label{HMm}
\end{equation}
where $\Delta_m=\delta_m - \sum_s\lambda_{m,s} \omega_{\text{vib},s}^{(m)}\big(b_{n,s}^{(m)} + b_{n,s}^{(m),\dagger}\big)$. $m$ denotes the aggregate and $s$ labels the degeneracy of vibrations. $\sigma_n^{(m),+}=|e_n^{(m)}\rangle\langle g_n^{(m)}|$ and $e_n^{(m)}$ represents the single electronically excited state at the $n$th chromophore. We define the phonon modes extending over the aggregate size, i.e, $b_{k,s}^{(m)} = N_m^{-1/2}\sum_{n=1}^{N_m} e^{-i k\cdot r_n} b_{n,s}^{(m)}$. From several studies, the absorption/fluorescence spectrum of aggregates shows a dense distribution of states attached to electronic excitations, as a result of the inhomogeneous line broadening characterized by a smooth spectral density of phonons \cite{Davydov_bok1971,Spano_JCP1991}. This indicates the significant contribution from the inter-molecular vibrations, and thus $b_{n,s}^{(m)}\approx N_m^{-1/2}\sum_k b_{k,s}^{(m)}$. Adding the coupling to cavity photons and rendering the bosonization of $H$, the full Hamiltonian forms with the bosonic operators $d_m, a$ for excitons and photons \cite{SM}. It thus reads $H=\sum_{m=1}^M h_m$ where
\begin{equation}
    \begin{split}
       h_m = & \delta_m d_m^{\dagger}d_m - \sum_{k,s} \frac{\lambda_{m,s}\omega_{\text{vib},s}^{(m)}}{\sqrt{N_m}} d_m^{\dagger}d_m \Big(b_{k,s}^{(m)} + b_{k,s}^{(m),\dagger}\Big) \\[0.15cm]
       & + \sum_{k,s}\omega_{\text{vib},s}^{(m)} b_{k,s}^{(m),\dagger}b_{k,s}^{(m)} + g_m\sqrt{N_m} \left(d_m^{\dagger} a + d_m a^{\dagger}\right).
    \end{split}
\label{Heff}
\end{equation}

{\it Full nonequilibrium dynamics}.--The exciton polaritons emerge from the cavity-aggregate interaction in Eq.(\ref{Heff}), i.e., $H_p = \sum_{m=1}^M \left[\delta_m d_m^{\dagger}d_m + g_m\sqrt{N_m} \left(d_m^{\dagger} a + d_m a^{\dagger}\right)\right] = \sum_{j=1}^{M+1}\varepsilon_j \eta_j^{\dagger}\eta_j$ where $[\eta_i,\eta_j^{\dagger}]=\delta_{ij}$ and $\eta_1,\ \eta_{M+1}$ annihilate particles at the lower polariton (LP) and upper polariton (UP) modes respectively, whereas $\eta_i;\ i=2,3,...,M$ annihilate at the $M-1$ dark modes. The interaction term in Eq.(\ref{Heff}) reads
\begin{equation}
    \begin{split}
       V_{\text{nr}}(t) = - \sum_{j,l=1}^{M+1}\sum_{m=1}^M & \sum_{k,s} \frac{f_{m,s}^{jl}}{\sqrt{2N_m}} \left[\eta_j^{\dagger}\eta_l b_{k,s}^{(m)} e^{i(\varepsilon_{jl}-\omega_{\text{vib},s}^{(m)})t} + \text{h.c.}\right] 
    \end{split}
\label{Vfa}
\end{equation}
invoking the rotating-wave approximation in the interactive picture and $f_{m,s}^{jl}=\sqrt{2}\lambda_{m,s}U_{mj}^* U_{ml}\omega_{\text{vib},s}^{(m)}$. $U$ is the unitary matrix diagonalizing $H_p$. Subject to an incoherent external pump $V_{\text{p}}(t) = \sum_{j=1}^{M+1}[\eta_j^{\dagger} F_j(t) e^{i\varepsilon_j t} + \eta_j F_j^*(t) e^{-i\varepsilon_j t}]$ with $\langle F_i^*(t)F_j(t')\rangle = \frac{r_j}{2}\delta_{ij}\delta(t-t')$, the full Hamiltonian is $V(t)=V_{\text{nr}}(t)+V_{\text{p}}(t)$. Averaging over the phonons and including the radiative loss, the coarse-grained equation for the density matrix is found
\begin{equation}
    \begin{split}
        \dot{\rho} = - \kappa \int_{-\infty}^t \int_{\tau'}^t \text{Tr}_{\text{E}} [V(t-\tau'),[V(\tau-\tau'), \varrho(\tau)]]\text{d}\tau \text{d}\tau'
    \end{split}
\label{rho}
\end{equation}
where $\varrho$ is the full density of polariton+phonon+radiation and $\rho=\text{Tr}_{\text{E}}(\varrho)$. $\kappa$ is the rate of scattering excitons by phonons. In what follows we will focus on identical aggregates $\delta_m=\delta,\ g_m=g,\ N_m=N$. Eq.(\ref{rho}) contains the rates of nonradiative relaxation and radiative loss,  which are thus $\chi_{jl}=\frac{\phi_{jl}}{MN} \bar{\chi}$ and $\gamma_j=MN \bar{\gamma}\varphi_j$ where
\begin{equation}
   \begin{split}
    \phi_{jl} = M\sum_{m=1}^M |U_{mj}|^2 |U_{ml}|^2,\quad \varphi_j = \frac{1}{M}\Big|\sum_{m=1}^M U_{mj}\Big|^2.
   \end{split}
\end{equation}

{\it Reduced dynamics of lower polaritons}.--Knowing the large mode density of the dark states, the reduced density matrix for the LP mode is defined 
\begin{equation}
    \sigma = \text{Tr}_{\text{ex}}\rho = \sum_{\{n_D, n_{M+1}\}}\langle n_2,...,n_{M+1}|\rho|n_2,...,n_{M+1}\rangle.
\label{rhopl}
\end{equation}
We define the operator ${\cal N}_j$ such that $\sigma {\cal N}_j = \text{Tr}_{\text{ex}}(\rho\eta_j^{\dagger}\eta_j)$ which turns out to be diagonal, possessing the representation
\begin{equation}
    \langle n'_1|{\cal N}_j|n_1\rangle = \langle {\cal N}_j\rangle_{n_1} \delta_{n'_1,n_1}
\label{Njma}
\end{equation}
where $\langle {\cal N}_j\rangle_{n_1}$ denotes the mean particle number at the $j$th mode, given $n_1$ particle at LP \cite{SM}. The calculations proceed as usual by inserting $\sigma$ and Eq.(\ref{Njma}) into Eq.(\ref{rho}), and we find the nonlinear quantum master equation (NQME)
\begin{widetext}
\begin{equation}
  \begin{split}
    \dot{\sigma} = \frac{1}{2}\Big[(S+1) \left(\eta_1\sigma\eta_1^{\dagger} - \sigma\eta_1^{\dagger}\eta_1\right) + S\left(\eta_1^{\dagger}\sigma\eta_1 - \sigma\eta_1\eta_1^{\dagger}\right)\Big] + \frac{\alpha}{2} \Big( \eta_1\sigma \mathscr{H}\eta_1^{\dagger} -  \sigma \mathscr{H}\eta_1^{\dagger}\eta_1 + \eta_1^{\dagger}\sigma \mathscr{K}\eta_1 - \sigma \mathscr{K}\eta_1\eta_1^{\dagger}\Big) + \text{h.c.}
  \end{split}
\label{qmel}
\end{equation}
\end{widetext}
with the heating and cooling operators $\mathscr{H}=\sum_{j=2}^{M+1}\mu_j\bar{n}_{j1} ({\cal N}_j+1),\ \mathscr{K}=\sum_{j=2}^{M+1}\mu_j (\bar{n}_{j1}+1) {\cal N}_j$, which depend on $\eta_1,\eta_1^{\dagger}$ \cite{SM}. $S=R_1+\bar{N}_1$ and $\bar{n}_{j1}$ is the mean thermal number of phonons. In present model, $\mathscr{H}$ and $\mathscr{K}$ are functions of $\sum_{j=2}^{M+1} {\cal N}_j=\Omega - \eta_1^{\dagger}\eta_1$, and $\Omega$ is the total particle number. 
$R_1=r_1/\gamma,\ \alpha=\chi/\gamma$.  $\chi=\chi_{j,1}=\phi_{j,1}\bar{\chi}/MN,\ (j=2,3,...,M)$. Essential parameters are $\bar{n}_L=[e^{(\tilde{\Omega}-\delta)/2T}-1]^{-1},\ \bar{n}'=(e^{\tilde{\Omega}/T}-1)^{-1}$ with the Rabi frequency $\tilde{\Omega} = \sqrt{\delta^2 + 4g^2 MN}$. $\bar{N}_j=(e^{\varepsilon_j/T}-1)^{-1}$.


It turns out that $[\Omega,H]=0$. $\Omega$ is thus a constant solely determined by pump and radiative loss. For various materials including molecules and semiconductors, the phonon-induced thermalization leads to a fast relaxation of excitons within 700fs, so that  $\bar{\chi}>\bar{\gamma}$. As such, $\dot{\Omega}=0$ so that its amount is predominately subject to the initial value created by the pump, yielding $\Omega \approx R + \bar{N}_1 + \bar{N}_{M+1}$ with $R=r/\gamma$ \cite{SM}.

{\it Out-of-equilibrium polariton condensation.}--For the mean number of LP, namely $\langle n_1\rangle$, the ansatz $\langle {\cal N}_{M+1}\rangle_{n_1}=\mathscr{A}_0$ is imposed, which takes the rational for the case $M\gg 1$. Proceeding via the NQME in Eq.(\ref{qmel}), we find the rate equation
\begin{equation}
    \begin{split}
        \langle \dot{n}_1\rangle = \alpha  (\Omega-C)\langle n_1\rangle - \alpha \langle n_1^2\rangle + \Lambda
    \end{split}
\label{n1da}
\end{equation}
with $C=1+\bar{n}_L M+\frac{1}{\alpha}+(1-\mu)\mathscr{A}_0 +\mu\bar{n}'$ 
and $\Lambda=R_1 + \alpha\left[(\bar{n}_L+1)\Omega - (\bar{n}_L-\mu\bar{n}'+1-\mu)\mathscr{A}_0\right]$. The LP mode experiences a gain due to the pump from all the modes and the loss energy via the terms $\alpha (\Omega-C) \langle n_1\rangle$ as well as $\alpha \langle n_1^2\rangle$, indicated by Eq.(\ref{n1da}). The condensation takes place once $\Omega>C$, giving the pumping threshold per aggregate ($w=R/M$)
\begin{equation}
   \begin{split}
     w_{\text{c}} = \bar{n}_L + \frac{1}{M\alpha} + \frac{1+\mu\bar{n}'+(1-\mu)\mathscr{A}_0}{M}
   \end{split}
\label{wc}
\end{equation}
as a reminiscence of a single-mode laser and a phonon condensation \cite{Frohlich_IJQC1968,Scully_PR1967}. Eq.(\ref{wc}) evidences the important role of dark states: a large mode density of dark states appreciably lowers the pumping threshold of polariton condensate, arising from the polariton nonlinearity. The $\mathscr{A}_0$-dependent term in Eq.(\ref{wc}) is a result from direct pumping from UP mode, which is less significant when $M\gg 1$.

\begin{figure}[t]
 \captionsetup{justification=raggedright,singlelinecheck=false}
\centering
\includegraphics[scale=0.215]{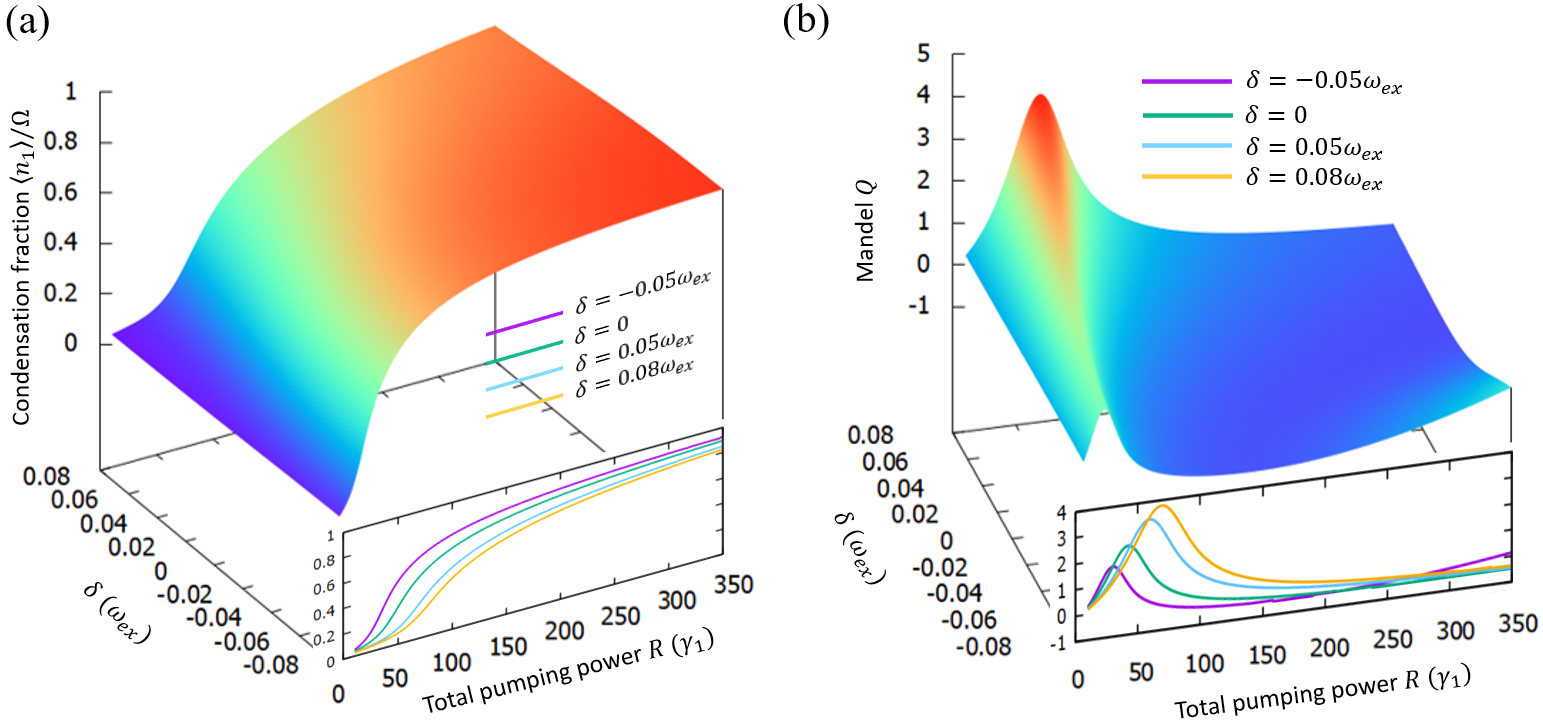}
\caption{(a) Condensate fraction and (b) Mandel parameter $Q$ of LP vs exciton-cavity detuning and pumping intensity. Parameters are $2g\sqrt{MN}=0.1\omega_{\text{ex}}$, $T=0.03\omega_{\text{ex}}$, $M=100$, $\alpha=0.07$, $R_1/R=0.02$ and $\mathscr{A}_0=6$, typically from organic molecules \cite{Forrest_NP2010,Xiong_NM2021}.}
\label{F2}
\end{figure}

The polariton condensation may be probed by the fluorescence spectrum, as illustrated in Fig.\ref{F1}(c). The condensation is evident by a sharp and intense peak at LP, followed by an asymmetric feature between UP and LP modes. The LP condensation is further characterized by the linewidth much narrower than the one for UP mode. This results from the coherence given by the off-diagonal components in Eq.(\ref{qmel}) that shows a much longer lifetime with the LP mode \cite{SM,Scully_PR1967}.

Nevertheless, the strong coupling and detuning between cavity and molecules may affect the threshold of LP condensation, evident by the $\bar{n}_L$ and $\bar{n}'$ in Eq.(\ref{wc}) and Fig.\ref{F2}(a) that elaborates the threshold shift as towards red detuning. Fig.\ref{F2}(a) also shows an over 80\% population at LP, with a strong pump. This may indicate unusual properties of quantum fluctuations, as will be entailed later.

{\it Polariton statistics.}--To access the quantum fluctuations of polariton condensates, we essentially calculate the steady-state statistics of the LP from Eq.(\ref{qmel}), solving for
\begin{equation}
    \begin{split}
        P_{n_1} = \frac{1}{Z_0} \left(\frac{{\cal Y}}{{\cal W}}\right)^{n_1} \frac{\left(\frac{{\cal X}}{{\cal Y}}-1\right)! \left(\frac{{\cal V}}{{\cal W}}-1-n_1\right)!}{\left(\frac{{\cal X}}{{\cal Y}}-1-n_1\right)! \left(\frac{{\cal V}}{{\cal W}}-1\right)!}
    \end{split}
\label{Pn1ss}
\end{equation}
where $Z_0=P_0^{-1}$ plays the role of partition function. ${\cal X} = R_1 + \bar{N}_1 + \alpha\big[(\bar{n}_L+1)(\Omega+1) + \left(\mu\bar{n}'-\bar{n}_L+\mu-1\right)\mathscr{A}_0\big]$, ${\cal Y} = \alpha(\bar{n}_L+1)$, ${\cal V} = R_1 + \bar{N}_1 + 1 + \alpha\big[\bar{n}_L(\Omega+M) + \left(\mu\bar{n}'-\bar{n}_L\right)(\mathscr{A}_0+1)\big]$ and ${\cal W} = \alpha\bar{n}_L$. The condensation suggests $P_{n_1-1}<P_{n_1}$ which indicates a bright accumulation, when $n_1<n_{\text{cr}}$, where $n_{\text{cr}}=M[w - \bar{n}_L - (M\alpha)^{-1} - (1+\mu\bar{n}'+(1-\mu)\mathscr{A}_0)/M]$. $n_{\text{cr}}>0$ gives the pumping threshold coinciding with the one from Eq.(\ref{wc}).

\begin{figure}[t]
 \captionsetup{justification=raggedright,singlelinecheck=false}
\centering
\includegraphics[scale=0.38]{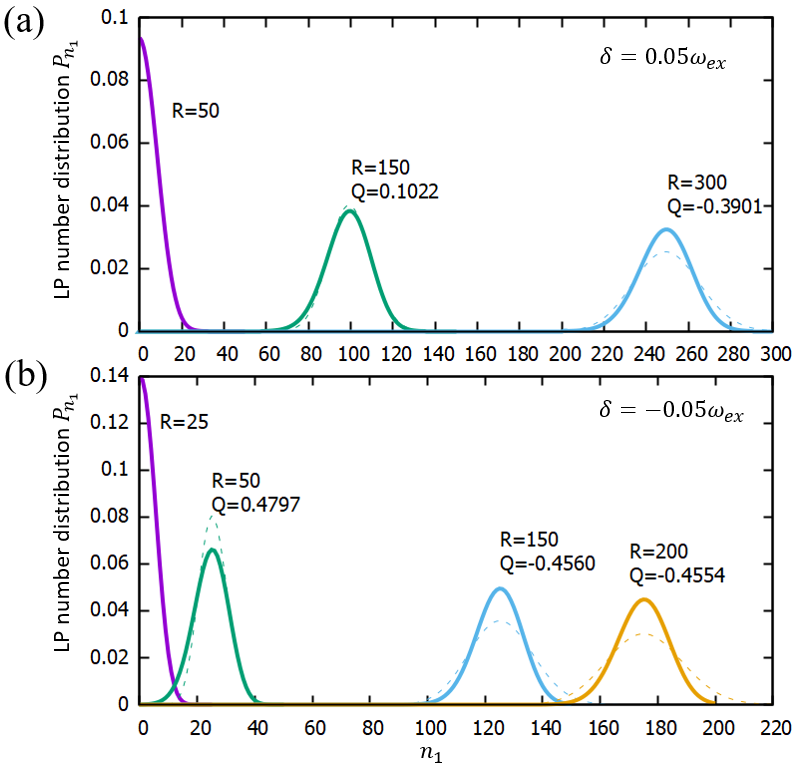}
\caption{Number statistics of LP mode for different detuning and pumping intensity. (a) Blue detuning $\delta=0.05\omega_{\text{ex}}$; (b) Red detuning $\delta=-0.05\omega_{\text{ex}}$. Purple lines reveal thermal distribution for pump below the threshold. Dashed lines represent the Poissonian distribution for different pumping intensities. Parameters are the same as Fig.\ref{F2}.}
\label{F3}
\end{figure}

Eq.(\ref{Pn1ss}) can access the fluctuations to all orders, e.g., $\langle n_1\rangle = ({\cal X}-{\cal V}-{\cal Y})/({\cal Y}-{\cal W})$ and higher order ones $\langle n_1^{\ell}\rangle$ which will be presented elsewhere. The Mandel parameter can be obtained thereby $Q=\langle \Delta n_1^2\rangle/\langle n_1\rangle -1$ \cite{SM}.


In a strong-pump limit, the condensate fraction $\eta=\langle n_1\rangle/\Omega$ and the $Q$ parameter approach asymptotically
\begin{subequations}
    \begin{align}
        & \eta \rightarrow 1 - \frac{1+\alpha \bar{n}_L M}{\alpha \Omega} \approx 1 - \frac{1}{w}\left(\frac{1}{M\alpha}+\bar{n}_L\right), \label{etas} \\[0.15cm]
        & Q \rightarrow \frac{R_1+1+\alpha\bar{n}_L M}{\alpha\Omega} - 1 \approx \frac{w_1}{\alpha w} + \frac{\bar{n}_L}{w} - 1. \label{Qasym}
    \end{align}
\end{subequations}
As shown, besides the total pumping intensity dependence, the polariton correlation relies on how strong the dark states are pumped whereas the mean LP number does not. Eq.(\ref{Qasym}) indicates a negative $Q$ with a large number of dark states being pumped, i.e., $M\gg 1$ yielding $Q\approx (M\alpha)^{-1} - 1$ given $\bar{n}_L\ll w$. For fewer dark states, however, $Q\approx 0.5 \alpha^{-1}-1$, which is positive normally; $Q=6.1$ using the parameters in Fig.\ref{F2}. 
Two parameter regimes for achieving nonclassical (sub-Poissonian) nature of the exciton-polariton condensation are identified: (1) $\chi>\gamma,\ \bar{n}_L\ll w_1$ when pumping the LP only; (2) $\chi M > \gamma,\ \bar{n}_L\ll w$ with the dark states pumped. This features $g^{(2)}(0)<1$ in the coincidence counting experiments, indicating the anti-bunching of polaritons.

Fig.\ref{F3} shows the LP statistics for different detuning and pumping intensity, elaborating the analytical solution in  Eq.(\ref{Pn1ss}). The density matrix formalism leads to the advantage of accessing the full spectrum of polariton fluctuations. We will be able to calculate the fluctuations to all orders, once knowing the density matrix elements. For blue detuning, as depicted in Fig.\ref{F3}(a), a thermal-like distribution is observed with pump intensity far below the threshold whereas a non-monotonic shape forms with pump intensity above the threshold. Such a transition has been measured in recent experiments of confined materials, manifesting the quantum fluctuation that deviates from the Bogoliubov theory \cite{Klaas_PRL2018,Pieczarka_NC2020}. It is worth noting a sub-Poissonian distribution for $Q<0$ at a strong pump. This suppresses the number fluctuation of LP condensate, making the LP mode stable, and therefore indicates a cooperative nature possessing an off-diagonal long-range order (ODLRO) \cite{Yang_RMP1962}. Fig.\ref{F3}(b) shows the LP statistics for red detuning, where the pumping power for nonclassical distribution is notably reduced when $\delta<0$. This can be seen from Eq.(\ref{Qasym}), in view of the lower $\bar{n}_L$. Nevertheless, a pump at rate of $r\approx 3$THz is estimated from Fig.\ref{F3}, for the nonclassical nature. This would be feasible because such a value is lower than the exciton frequency $\sim 2$eV \cite{Forrest_NP2010,Xiong_NM2021,Lidzey_Nature1998}. 

More information about LP statistics is offered by the $Q$ parameter, as shown in Fig.\ref{F2}(b). The $Q$ value is sensitive to both the detuning and pumping power. The ridge at low pumping power indicates a weak bunching of polaritons when approaching $\delta<0$. At high pumping power, the $Q$ drops sharply, yielding a widen region for sub-Poissonian distribution. This may indicate a broad and robust parameter space to achieve highly-quantum light emission.

\begin{figure}[t]
 \captionsetup{justification=raggedright,singlelinecheck=false}
\centering
\includegraphics[scale=0.26]{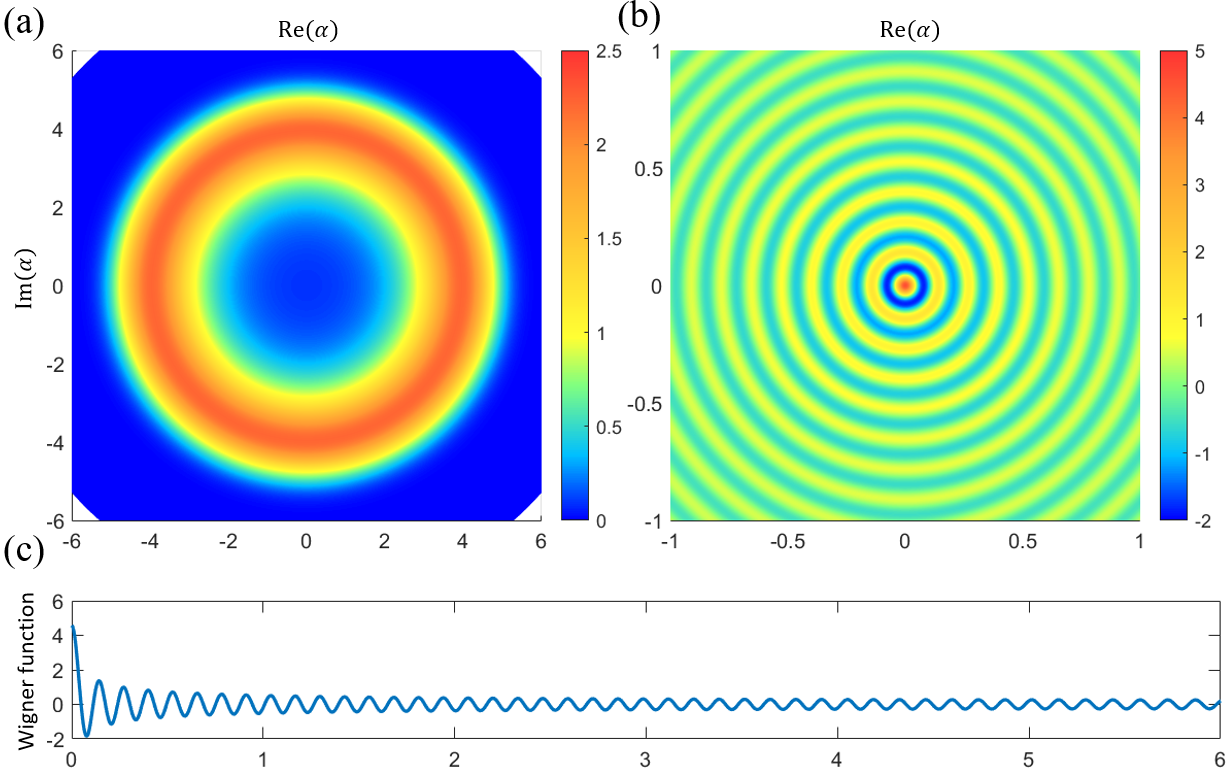}
\caption{Wigner function of LP mode for different pumping intensities. (a) $R=40$; (b) $R=150$ with piece-wise negativity; (c) Radial projection of (b), provided the modulus dependence in Eq.(\ref{W}). Other parameters are the same as Fig.\ref{F2}.}
\label{F4}
\end{figure}

The phase transition to LP condensation can be elaborated in phase space, using the Wigner function
\begin{equation}
    W(\alpha) = \frac{2}{\pi} e^{-2|\alpha|^2}\sum_{n_1=0}^{\Omega} (-1)^{n_1} L_{n_1}(4|\alpha|^2) \sigma_{n_1,n_1}
\label{W}
\end{equation}
where $L_n(x)$ is the Laguerre polynomial of $n$th order. Fig.\ref{F4} shows the nonclassical nature of the LP condensation evident by the piece-wise negativity at $Q<0$, whereas the condensation with $Q>0$ illustrates a single-ring geometry that behaves as a laser.

{\it Photon counting statistics}.--Since the phonon-induced relaxation is faster than the polartion radiation, the emitted photons from the condensates can be detected \cite{Lagoudakis_PRL2022}. 
Considering the emission off the cavity axis, the polariton-photon interaction is $V(t)\propto \eta_1^{\dagger}a e^{i(\varepsilon_1+v_c-v)t} + \text{h.c.}$ provided that the LP mode is spectrally resolved. 
Microscopically, the LP emits photons in a random fashion such that the emission could take place at any time. This makes the timed coarse graining essential for the equation of motion for the joint density matrix of LP + photons. 
Some algebra gives the joint population \cite{photon}
\begin{equation}
    \begin{split}
        \dot{P}_{m,n_1} = - (m+1)(n_1 - m) P_{m,n_1} + m (n_1-m+1) P_{m-1,n_1}
    \end{split}
\label{Pm}
\end{equation}
with $m\le n_1$, where $m$ denotes the photon number. In a long-time limit, $\dot{P}_{m,n_1}=0$ yielding the solution $P_{m,n_1}=P_{n_1}\delta_{m,n_1}$ given $P_{n_1}$ by Eq.(\ref{Pn1ss}). The photon-number distribution reads
\begin{equation}
    \mathscr{P}_m = \sum_{n_1} P_{m,n_1} = P_m.
\end{equation}
Therefore the emission statistics is dictated by the number distribution of the condensate \cite{Expl}.

{\it Conclusion and remarks}.--To show the feasibility of our model in reality, proper candidates of organic molecules would be the J-aggregates and anthracenes. Their large dipole moments result in a strong coupling to cavity photons, yielding the Rabi splitting $\tilde{\Omega}\approx 80-150$meV \cite{Forrest_NP2010,Xiong_NM2021}. Noting the room temperature $= 26$meV, $\tilde{\Omega}/T \approx 3-6$ close to the parameter regime in Fig.\ref{F2}. Moreover, both of the two systems possess the exciton binding energy larger than traditional inorganic semiconductors, which makes them stable at room temperature and robust against dissociation in external fields.

So far, we have developed a full quantum theory for the exciton-polariton condensate with molecules, demonstrating a nonequilibrium phase transition. The microscopic model was solved analytically, going beyond the mean-field description that cannot access the full statistics of condensated polaritons. Notably, the nonclassical property is predicted for the exciton-polariton condensates, beyond the understanding at ODLRO level. Our results identify the role of dark states with a high density important for the condensates. 
Nevertheless, a nonlinear master equation having the form of Eq.(\ref{qmel}) would be generic for any condensation mechanism. It thus provides a comprehensive approach for calculating the nonequilibrium dynamics and fluctuation of cavity polaritons. This offers new insights for molecular polaritons perspective for room-temperature polaritonics and quantum-light sources.

\vspace{0.15cm}
We thank Zhe-Yu Jeff Ou from City University of Hong Kong and Renbao Liu from Chinese University of Hong Kong, for their instructive discussions. Z.D.Z. and S.Z. gratefully acknowledge the support of Early Career Scheme from Hong Kong Research Grants Council (No. 21302721) and National Science Foundation of China (No. 12104380). D.L. gratefully acknowledges the support of AoE grant from Hong Kong Research Grants Council (No. AoE/P-701/20).


\begin{thebibliography}{}
\bibitem{photon}
The timed coarse-grained equation leads to the diagonal elements $\dot{P}_{n,m} = - \gamma (m+1)n P_{n,m} + \gamma m (n+1)  P_{n+1,m-1}$ with a conserved quantity $n_1=n+m$, keeping the most significant terms. 
This is due to the fact that the emitted photons will be captured by detectors rather than bouncing back. $n$ and $m$ denote the numbers of LP and emitted photons, respectively. One can define the photon population given $n_1$ particles at LP mode, i.e., $P_{m,n_1}=P_{n,m}$. This results in the equation for $P_{m,n_1}$. Noticing the elementary process such that a LP quanta is converted to a photon, the $n_1$ must be the LP number prior to the emission.
\bibitem{Expl}
General to speak, the lasing photons from exciton-polariton condensates are detected within a finite timescale during the emission. The full time-dependent solution to the equation for $P_{m,n_1}$ is thus necessary, yielding the photon-number distribution $\mathscr{P}_m(t)$. The exact solution can be obtained in a math rigor and will be presented elsewhere, to avoid redundancy.
\end{thebibliography}
\end{document}